\documentclass[preprint,final,12pt]{elsarticle}
\usepackage{textcomp}
\usepackage{amssymb}
\usepackage{mathtools}
\usepackage{amsmath}
\usepackage{type1cm}
\usepackage{geometry}
\usepackage{color}
\usepackage{epsfig}
\usepackage{latexsym}
\usepackage{setspace}
\usepackage{hyperref}
\usepackage[all]{hypcap}
\usepackage{graphicx}
\usepackage{epstopdf}
\usepackage{float}
\usepackage{subfig}
\usepackage{lineno}
\usepackage{textcomp,gensymb}

\journal{Astroparticle Physics Journal \date{\today}}

\begin{document}

\begin{frontmatter}
  
\title{The Knee and the Second Knee of the Cosmic-Ray Energy Spectrum}

\author{T. Abu-Zayyad}
\corref{correspondingauthor}
\cortext[correspondingauthor]{Corresponding author\\ E-mail: tareq@cosmic.utah.edu}
%\author{R. Cady}
\author{D. Ivanov},
\author{C.C.H. Jui}
\author{J.H. Kim}
\author{J.N. Matthews}
\author{J.D. Smith}
\author{S.B.~Thomas}
\author{G.B. Thomson}
\author{Z. Zundel\\High Energy Astrophysics Institute and Department of
  Physics and Astronomy, University of Utah, Salt Lake City, Utah, USA}

\begin{abstract}
The cosmic ray flux measured by the Telescope Array Low Energy
Extension (TALE) exhibits three spectral features: the knee, the dip
in the $10^{16}$ eV decade, and the second knee. Here the spectrum has
been measured for the first time using fluorescence telescopes,
which provide a calorimetric, model-independent result. The spectrum
appears to be a rigidity-dependent cutoff sequence, where the knee is
made by the hydrogen and helium portions of the composition, the dip
comes from the reduction in composition from helium to metals, the
rise to the second knee occurs due to intermediate range nuclei,
and the second knee is the iron knee.
\end{abstract}

\end{frontmatter}

\section{Introduction}

The spectrum \cite{talespec} recently measured by the TALE
fluorescence detector of the Telescope Array experiment covers the
energy range, $10^{15.3}\,\mathrm{eV} < E < 10^{18.3}\,\mathrm{eV}$,
and represents the first time that fluorescence telescopes have
observed this low in energy. In this energy range there are three
spectral features. In the TALE spectrum the knee appears as a broad
maximum centered at $10^{15.6}$ eV, there is a broad dip centered at
$10^{16.2}$ eV, and the second knee occurs at $10^{17.04}$ eV. The
energy scale of TALE is the same as that of the entire TA experiment,
and is consistent with the energy of the GZK cutoff \cite{greisen,zk}
which is observed at $10^{19.75}$ eV \cite{sdspec_5yr}. The energy
resolution of the TALE fluorescence detector is about $15\%$ and is
constant as a function of energy. This represents the first time that
this energy range has been observed calorimetrically, with excellent
energy resolution.

The Kascade experiment \cite{kascade} was the first to see the
systematic change in composition as the energy increases above the
knee of the spectrum. This observation was made using ground array
detectors sensitive to the muonic and electromagnetic components of
cosmic ray air showers. Although the specific changes in composition
they described were model-dependent, the interpretation was widely
accepted as representing a rigidity-dependent cutoff sequence at the
end of the galactic cosmic ray spectrum. Their interpretation was that
the hydrogen knee was at about $10^{15.5}$ eV.  For a
rigidity-dependent cutoff sequence, this would put the iron knee at
$10^{16.9}$ eV.  However the Kascade experiment could not make
reliable measurements this high in energy.

\begin{figure*}[htb]
  \centering
  \subfloat[]{\includegraphics[width=0.5\textwidth]{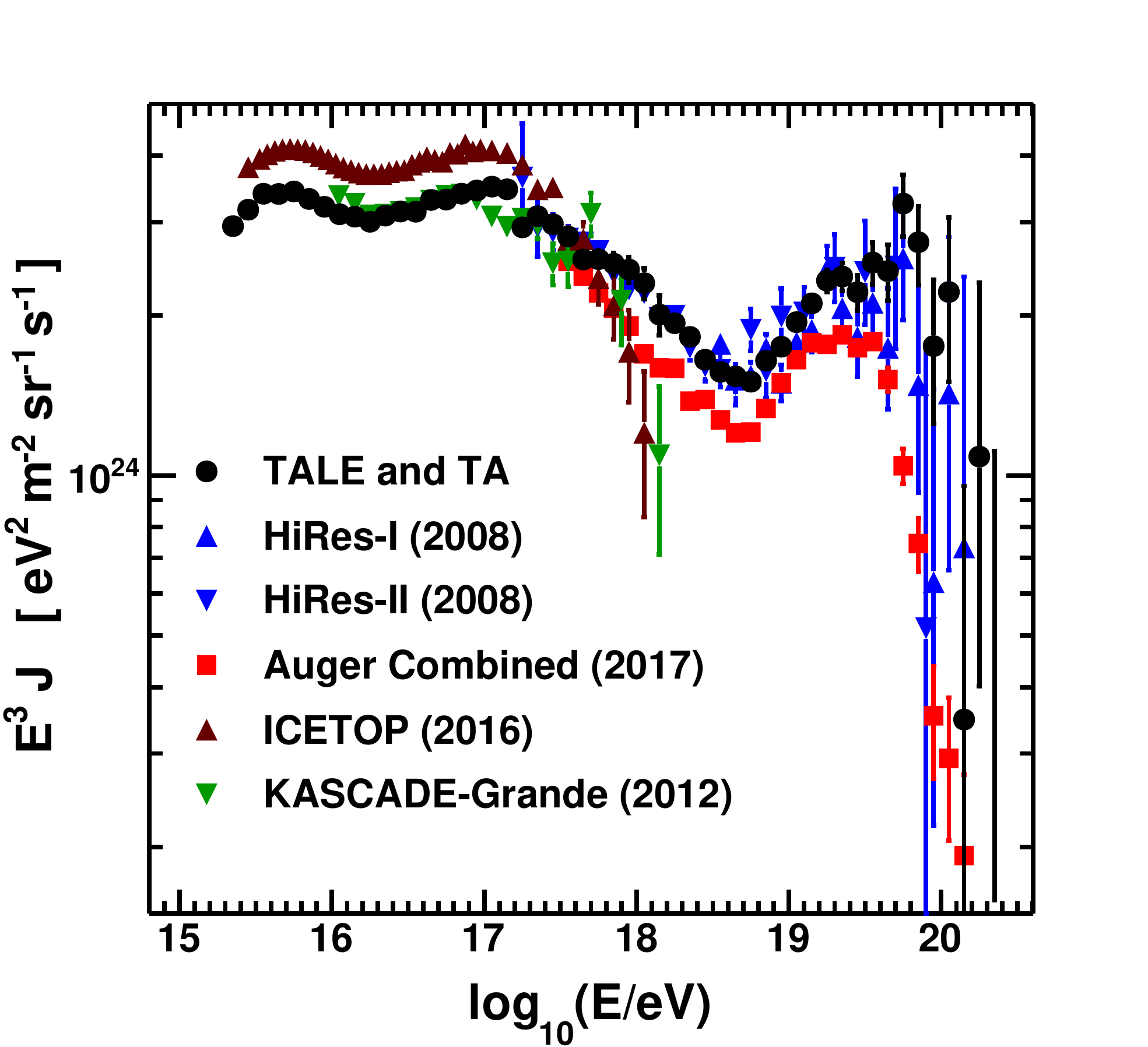}\label{fig:tatale_and_others_nominal}}
  \subfloat[]{\includegraphics[width=0.5\textwidth]{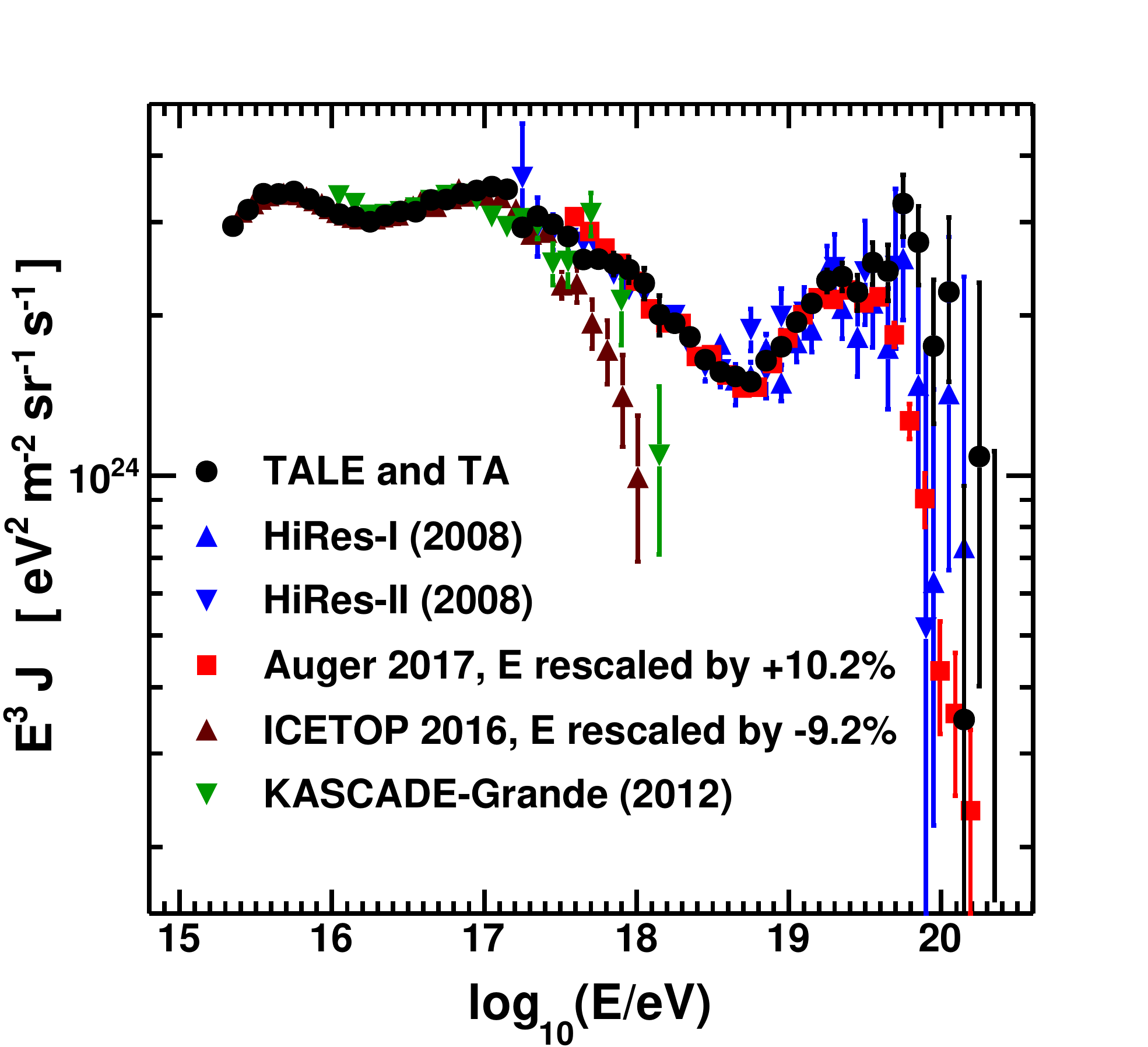}\label{fig:tatale_and_others_rescaled}}
  \caption{ \protect\subref{fig:tatale_and_others_nominal} Cosmic Ray
    spectrum measured by TALE, TA, Auger, HiRes, IceTop, and
    Kascade-Grande. The TALE and TA spectrum is obtained by combining the
    TALE ~\cite{talespec} and TA surface detector
    ~\cite{ta_spectrum_icrc2017} spectra.
    \protect\subref{fig:tatale_and_others_rescaled} The same plot, but
    the energy scale of Auger is raised by $10.2\%$, and that of IceTop
    is lowered by $9.2\%$. }
  \label{fig:tatale_and_others}
\end{figure*}

Subsequent experiments have seen the three spectral features, with
different energy scales.  Figure \ref{fig:tatale_and_others_nominal}
shows the spectra of TALE ~\cite{talespec}, Telescope Array (TA)
~\cite{ta_spectrum_icrc2017}, HiRes ~\cite{hires_spectrum}, Pierre
Auger Observatory ~\cite{auger_spectrum_icrc2017}, IceTop
~\cite{icetop}, and Kascade-Grande ~\cite{kascade-grande}. Some of the
differences in the spectra seem to be due to different energy scales
of experiments. Figure \ref{fig:tatale_and_others_rescaled} shows the
same spectra, but with the energy scale of IceTop reduced by $9.2\%$
and the Auger energy scale increased by $10.2\%$. Here the IceTop and
TALE spectra agree up to about $10^{17.5}$ eV, the TA and Auger
spectra agree up to about $10^{19.5}$ eV. TALE, HiRes, and
Kascade-Grande agree without energy scale adjustment. Thus there is a
broad consensus that there exist features called the knee, the dip in
the $10^{16}$ eV decade, and the second knee, and that we have a good
idea of the energies at which the features occur.

In what follows we argue that the second knee is the iron knee, and
the spectrum is consistent with a rigidity-dependent cutoff
sequence. First we argue on general grounds, then we present three
models that illustrate fits to the spectrum.

\section{The TALE Experiment}

The TALE experiment consists of two sets of detectors of cosmic rays:
a fluorescence detector (FD) consisting of 10 telescopes that look
high in the sky (at elevations from $31\degree$ to $59\degree$), and a
surface detector consisting of 103 scintillation counters forming an
infill array in front of the FD. The TALE spectrum in reference
\cite{talespec} was based on observations made with the FD
alone. Events seen by the FD below about $10^{17}$ eV were dominated
by Cherenkov light from cosmic ray shower particles, but above
$10^{17.5}$ eV, fluorescence light was the largest contribution to the
signal. In the intermediate energy range, a mixture of Cherenkov and
fluorescence light pertained. The energy resolution of the TALE FD is
about $15\%$, independent of energy, and the spectrum measured is
independent of models.

\section{The Rigidity-dependent Cutoff Sequence}

The basic idea of a rigidity dependent cutoff sequence is that in a
cosmic ray accelerator a moving magnetic field accelerates
particles. The maximum energy achieved by the accelerator depends upon
the strength of the magnetic field, its speed of movement, and the
time duration that particles are in contact with it. In this situation
the maximum energy of nuclei will be proportional to their charge;
i.e., the cutoff rigidity (energy/charge) is the same for all nuclear
species. Thus if the maximum energy of hydrogen nuclei is $E$, for
helium it is $2\times E$, carbon is $6\times E$, etc. Practically
speaking, the sequence ends with iron, at an energy of $26\times E$.
If we start with the interpretation of the Kascade experiment as a
guide, but using the TALE spectrum energy scale, the second knee can be
identified as the iron knee, at $10^{17.04}$ eV. Dividing by 26, the
proton maximum of the spectrum would be $10^{15.6}$ eV, and the helium
maximum would be at $10^{15.9}$ eV. The broad maximum of the knee is
then identified as the result of H and He coming to their maximum
energies.  The abundance of metals is considerably lower than H and
He, so a dip in the spectrum should occur at higher energies.  In all
cosmic scenarios, the abundance of Li, Be, and B is very low, which
enhances the dip. Although the CNO group is more abundant, it is much
less so than H and He. Thus, there should be a broad minimum in the
spectrum of a rigidity-dependent cutoff sequence, with a rise near the
carbon location of $10^{16.4}$ eV. This is what is seen in the TALE
spectrum. Due to intermediate weight nuclei the spectrum rises (on an
$E^3J$ plot), then peaks at Fe.

\section{The Extragalactic Contribution}

One detail that must be taken into account is the low energy end of
the extragalactic cosmic ray flux. All experiments with fluorescence
detectors that can measure the depth of shower maxima indicate that
between $10^{18.0}$ and $10^{18.5}$ eV the composition is very light,
and probably protonic \cite{hires-comp} \cite{tamd-comp}
\cite{brlr-comp} \cite{pao-comp}. If these cosmic rays originated
within our galaxy, there would be considerable anisotropy in their
arrival directions, but this is not seen either in the northern
\cite{ta_EeV_protons} or southern \cite{auger-anisot}
hemispheres. Hence, one expects these protons to be of extragalactic
origin. This is a general result. Using galactic magnetic field
models, one limit \cite{ta_EeV_protons} has been put that (at $95\%$
confidence level) $<1.6\%$ of cosmic rays are of galactic origin. By
extension, an extragalactic flux of light composition must extend down
to energies lower than $10^{18}$ eV.  One can estimate the
contribution of extragalactic protons in the $10^{17}$ eV decade using
measurements of Xmax, the depth of shower maximum. The HiRes-MIA
\cite{hrmia_comp} and Auger \cite{auger_xmax_icrc2017} measurements
indicate that at $10^{17}$ eV the mean of Xmax is midway between what
is expected for hydrogen and iron. Perhaps half of cosmic rays at this
energy are extragalactic protons.

\section{\texorpdfstring{Three Specific Models of the $10^{15} - 10^{18}$ Decades}{Three Specific Models of the 10**15 - 10**18 Decades}}

\begin{figure}[htb]
\centering
\includegraphics[width=0.7\textwidth]{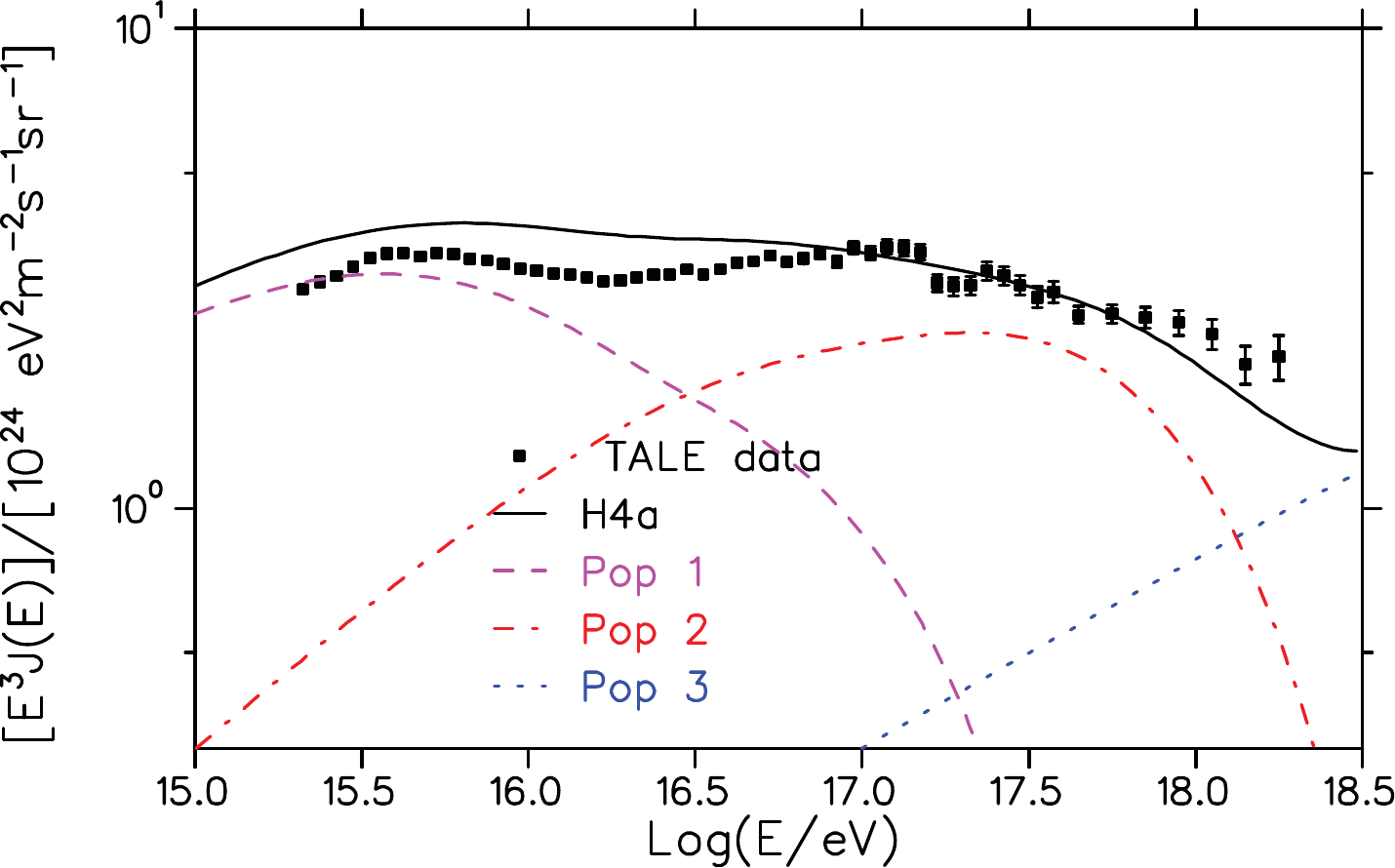}
\caption{Cosmic Ray spectrum measured by TALE between $10^{15.3}~eV$ and
$10^{18.3}$ eV, overlaid with the H4A model described in this
section.} 
\label{fig:h4a}
\end{figure}

\begin{figure}[htb]
\centering
\includegraphics[width=0.7\textwidth]{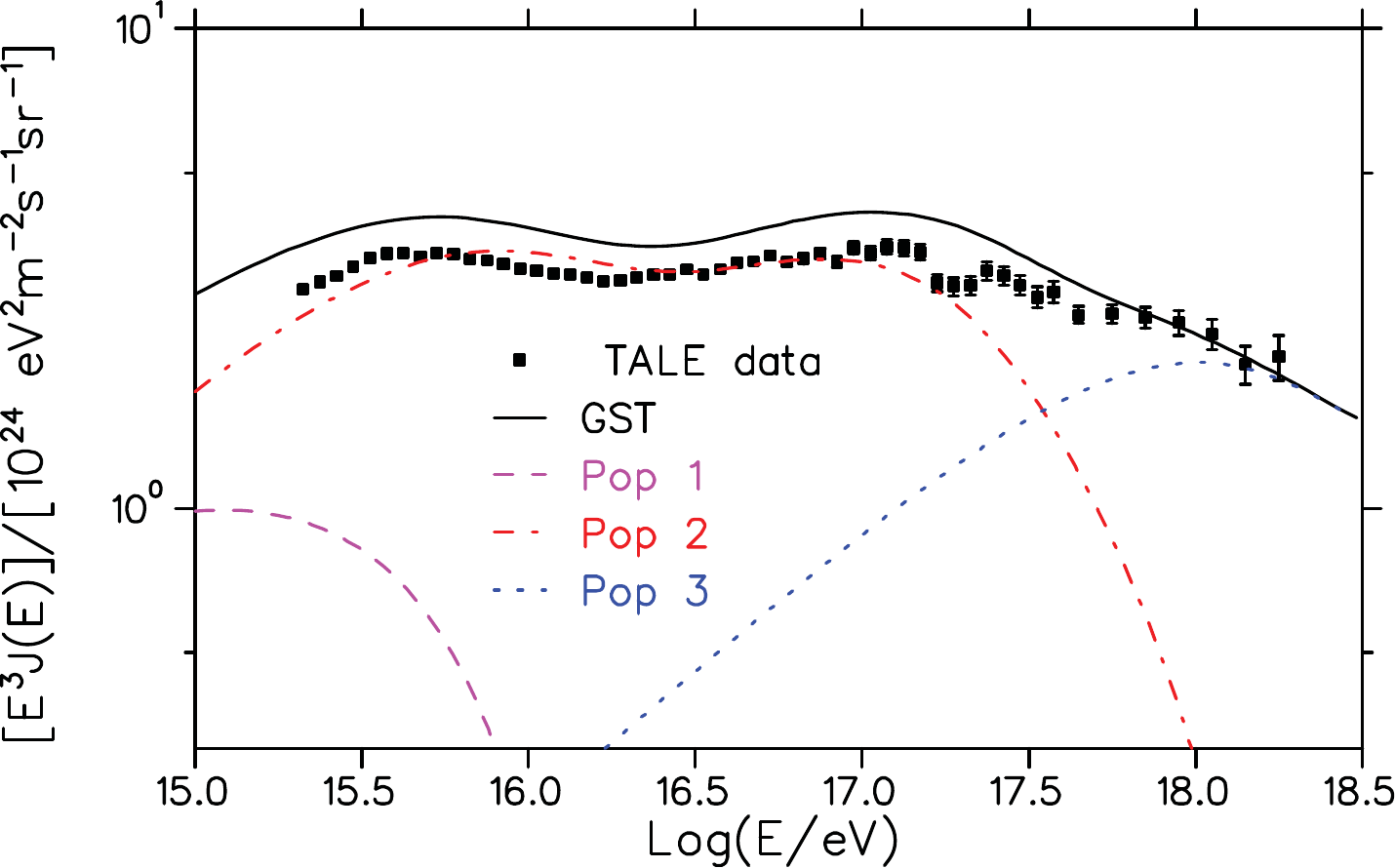}
\caption{Cosmic Ray spectrum measured by TALE between $10^{15.3}~eV$ and
$10^{18.3}$ eV, overlaid with the GST model described in this
section.} 
\label{fig:gst}
\end{figure}

Here we wish to present three models of the rigidity-dependent cutoff
sequence in comparison with the TALE spectrum: the H4a model by
T. Gaisser \cite{h4a_comp}, a model by T. Gaisser, T. Stanev, and
S. Tilav (GST) \cite{gst}, and a model taken from direct measurements
of composition in the $10^{13}$ eV decade \cite{PDG2016} and
extrapolated to higher energies by 2.5 decades. The H4a and GST models
include an extragalactic component, but in the direct measurement
model we have supplied an estimate of the extragalactic flux.  The H4a
model assigns the features of the spectrum to three populations of
cosmic rays, two of galactic origin and the third of extragalactic
origin. The first population forms the knee, and the second population
makes the second knee. The dip in the $10^{16}$ eV decade seems to be
poorly expressed in this model. Figure \ref{fig:h4a} shows the
comparison of the H4a model to the TALE spectrum. Figure \ref{fig:gst}
shows the comparison of the GST model (which also uses three
populations) to the TALE spectrum.  The GST model might have a higher
energy scale than the TALE data.

In the third model, we attempt to recreate the observed spectrum from
extrapolating the measured composition at lower energies, and applying
a rigidity-dependent cutoff for the termination of the dominant
galactic components.  We also assume a very simple phenomenological
model for an extragalactic component at higher energies.  Hence this
is a model of 2 populations, one galactic and one extragalactic.

The known lower-energy data is obtained by extrapolation from the
Particle Data Group's compilation given in Figure 29.1 of the Particle
Data Book \cite{PDG2016}.  Assuming the 11 nuclei displayed are the
dominant species, we interpolated relative abundances at $10^{13}$~eV.
These values (normalized to one) are shown in Table~\ref{tab:abun13}.
Other species are neglected in our model.  In this energy regime, all
species shown in Figure 29.1 appear to follow a common power law
$E^{-\alpha}$ with an index of $\alpha=2.8$.  We assume this value in
our model.

\begin{table}[htb]
\centering
\caption{Normalized abundances, at $10^{13}$~eV,
of the 11 nuclei from the Particle Data Book}
\label{tab:abun13}
\begin{tabular}{|l|r|c|}
\hline
Element & Z & fraction at $10^{13}$~eV \\ \hline
hydrogen & 1 & 0.3019 \\
helium & 2 & 0.4104 \\
carbon & 6 & 0.0388 \\
oxygen & 8 & 0.0745 \\
neon & 10 & 0.0153 \\
magnesium & 12 & 0.0293 \\
silicon & 14 & 0.0308 \\ 
sulfur & 16 & 0.0082 \\
argon & 18 & 0.0043 \\
calcium & 20 & 0.0070 \\
iron & 16 & 0.0800 \\
\hline
\end{tabular}
\end{table}

We attempt to explain the occurrence of the knee and the second knee
as the result of terminations in the acceleration of the 11 nuclei.
The flux observed on Earth could be dominated by one local and recent
galactic source, or a class of sources.  In such a scenario, if
protons are accelerated up to a cut-off energy $E_p$, then the cut-off
energies for heavier species should be given by $ZE_{p}$.  However, an
abrupt, step-function cut-off is clearly unphysical.  Cut-offs are
often modeled as an exponential decay above the break, resulting from,
for instance Bohm diffusion. However, it is possible that there are
additional, weaker sources that extend to higher energies still.  We
therefore model the break by a broken power law where the slope of the
spectrum in each case changes from $\alpha=2.8$ to a larger (steeper)
value $\beta$.

The extra-galactic component is assumed to follow a simple
power law with a log-exponential cut-off represented by
\begin{equation}
\log_{10}[J_{XG}(E)] = B + (1-e^{\frac{c-x}{d}}) - \gamma{x}
\end{equation}
where $x=\log_{10}{E}$, and the $-\gamma{x}$ term represents the
extrapolation of the piece-wise power-law spectrum previously seen
below the ankle feature. The low-energy cut-off for the extra-galactic
component is assumed to occur at $10^{c}$~eV, smoothed by a broadening
of the cut-off with width $d$.  This form for the cut-off was chosen
for its simplicity.

\begin{figure}[htb]
\centering
\includegraphics[width=\columnwidth]{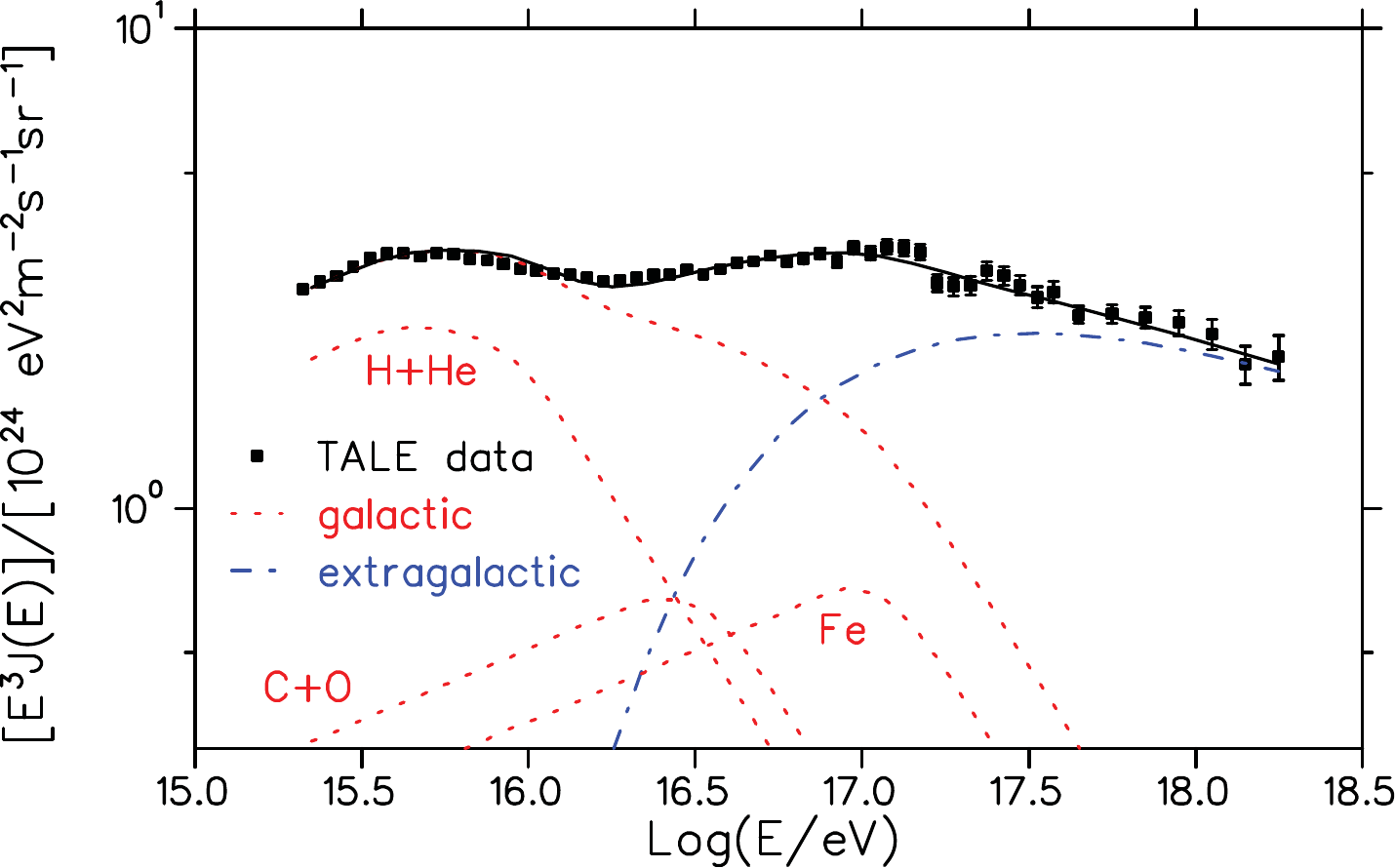}
\caption{Cosmic Ray spectrum measured by TALE between $10^{15.3}~eV$ and
$10^{18.3}$ eV, overlaid with the phenomenological model described in this
section.  The galactic component, extrapolated from Figure 29.1 of the
Particle Data Book, is shown by the red dot-dashed curve.  The proton
acceleration cut-off is placed at $10^{15.6}$~eV.  Pre-break post-break
power indices of $\alpha=2.80$ and $\beta=4.20$ are assumed, respectively,
for all species. The galactic component is described by a power law of
index $\gamma=3.24$, with a cut-off at
$10^{16.4}$~eV and width
of $c=0.55$.}
\label{fig:spec}
\end{figure}

Figure~\ref{fig:spec} shows the TALE energy spectrum overlaid with the
simple model described in the previous paragraphs.
Table~\ref{tab:params} lists the parameter values that combine to give
a total flux that gives good agreement to the shape of TALE data.  The
group contributions of H+He, C+O, and Fe are shown separately in
addition to the total galactic component.

\begin{table}[htb]
\centering
\caption{Model parameters used to calculate the TALE spectrum}
\label{tab:params}
\begin{tabular}{|c|c|l|}
\hline
symbol & value & explanation \\ \hline
$\alpha$ & 2.80 & galactic spectral power index below cut-off energy \\
$\beta$ & 4.20 & galactic spectral power index above cut-off energy \\
$E_p$ & $10^{15.60}$~eV & cut-off energy for protons \\
$\gamma$ & 3.24 & extra-galactic spectral index  \\
$d$ & 0.55 & width of extra-galactic cut-off \\
$c$ & 16.4 & log-Energy of cut-off of extra-galactic flux \\
\hline
\end{tabular}
\end{table}

%\begin{figure*}[htb]
%\centering
%\subfloat[]{\includegraphics[width=0.5\textwidth]{hi03.pdf}\label{fig:spec_pdg_comp}}
%\subfloat[]{\includegraphics[width=0.5\textwidth]{hi03.pdf}\label{fig:spec_pdg_comp_p_he_reversed}}
%\caption{\protect\subref{fig:spec_pdg_comp} Cosmic Ray spectrum
%  measured by TALE between $10^{15.3}~eV$ and $10^{18.3}$ eV, overlaid
%  with the phenomenological model described in this section.  The
%  galactic component, extrapolated from Figure 29.1 of the Particle
%  Data Book, is shown by the red dot-dashed curve.  The proton
%  acceleration cut-off is placed at $10^{15.6}$~eV.  Pre-break
%  post-break power indices of $\alpha=2.80$ and $\beta=4.20$ are
%  assumed, respectively, for all species. The galactic component is
%  described by a power law of index $\gamma=3.24$, with a cut-off at
%  $10^{17.0}$~eV and a log-exponential slope of
%  $c=0.30$. \protect\subref{fig:spec_pdg_comp_p_he_reversed} The same
%  plot but where the hydrogen and helium abundances at $10^{13}$~eV
%  are reversed}
%\label{fig:spec}
%\end{figure*}

\section{Conclusions}

The TALE spectrum shows three spectral features which resemble the
effects of a rigidity-dependent cutoff sequence, which could occur at
the high energy end of the galactic cosmic ray spectrum. The general
features of such a sequence are: a broad knee made from H and He, a
drop in spectrum (a dip) caused by the much lower abundance of metals,
and a rise to an iron knee marking the very end of the galactic
spectrum. Identifying the second knee seen in the TALE spectrum with
the iron knee, the energies of the features seen correspond closely
with this general picture.  One can also compare specific models of
the rigidity-dependence sequence to the TALE spectrum. The H4a model
uses three populations to form the knee, the second knee, and an
extragalactic population, but overshoots the dip.  The GST model has a
stronger dip. A model using the PDG compendium as a starting point can
form an acceptable fit to the spectrum when one adds an extragalactic
component.  In all these cases we identify the second knee at
$10^{17.04}$ eV as the rigidity-dependent cutoff of iron.

\section{Acknowledgements}

The authors wish to thank the members of the University of Utah Cosmic
Ray Group for many interesting discussions, and the U.S. National
Science Foundation for its awards PHY-0601915, PHY-1404495,
PHY-1404502, and PHY-1607727.

%\newpage
\bibliographystyle{elsarticle-num}
%%\bibliography{tlinterp}

\end{document}